\documentclass[twocolumn,aps,prb,tightenlines,amsmath,amssymb,superscriptaddress]{revtex4}
\usepackage{graphicx}
\usepackage{dcolumn}
\usepackage{colordvi}
\usepackage{color}

\begin{document}
\hyphenation{spectro-meter}
\hyphenation{micro-scopy}
\hyphenation{image}
\hyphenation{Na-tu-rally}
\title{Time-resolved Kerr rotation spectroscopy of valley dynamics in single-layer MoS$_2$}
\author{G.\ Plechinger}
\affiliation{Institut f\"ur Experimentelle und Angewandte Physik,
Universit\"at Regensburg, D-93040 Regensburg, Germany}
\author{P.\ Nagler}
\affiliation{Institut f\"ur Experimentelle und Angewandte Physik,
Universit\"at Regensburg, D-93040 Regensburg, Germany}
\author{C.\ Sch\"uller}
\affiliation{Institut f\"ur Experimentelle und Angewandte Physik,
Universit\"at Regensburg, D-93040 Regensburg, Germany}
\date{\today}
\author{T.\ Korn}
\email{tobias.korn@physik.uni-regensburg.de}
\affiliation{Institut
f\"ur Experimentelle und Angewandte Physik, Universit\"at
Regensburg, D-93040 Regensburg, Germany}
\begin{abstract}
Single-layer MoS$_2$ and similar dichalcogenides are direct-gap semiconductors with a peculiar band structure: the direct gap is situated at the K$^+$ and K$^-$ points in the Brillouin zone, with a large valence-band spin splitting. Optical selection rules allow for valley-selective interband excitation using near-resonant, circularly polarized excitation.  Here, we present time-resolved pump-probe experiments in which we study the carrier and valley dynamics in a mechanically exfoliated single-layer MoS$_2$ flake at low temperatures. Under resonant excitation conditions, we find that the valley lifetime exceeds the photocarrier lifetime, indicating the creation of a resident valley polarization. For highly nonresonant excitation, the valley polarization decays within  the photocarrier lifetime.
\end{abstract}
\maketitle
In recent years, two-dimensional (2D) crystal structures such as graphene have attracted a lot of scientific interest. Besides graphene and its bulk crystal graphite, there is a large number of layered crystal structures, in which covalently bonded 2D sheets are weakly coupled in the bulk crystal by van der Waals forces. Single 2D layers can be isolated from many of these bulk crystals by mechanical exfoliation. Therefore, a variety of different material classes is readily available as a 2D sheet~\cite{Novoselov26072005}, including large-gap insulators, superconductors, and semiconductors. MoS$_2$ and related transition-metal dichalcogenides are among the most promising systems: while they are indirect-gap semiconductors in the bulk, a transition to a direct band gap situated at the  K$^+$/K$^-$ points in the Brillouin zone occurs in single layers~\cite{Eriksson09,Heine_PRB11}, and a pronounced increase of photoluminescence (PL) yield is observed as the number of layers is reduced~\cite{Heinz_PRL10,Splen_Nano10}. Additionally, there is a large spin splitting of the valence band (about 150~meV for MoS$_2$)  at the K points. Optical selection rules for interband transitions allow for valley-selective excitation of electron-hole pairs via circularly polarized light~\cite{Xiao12}, and the optically oriented valley polarization can be read out using helicity-resolved photoluminescence (PL)~\cite{Yao12,Heinz12,Zeng_WS2_12, Marie_Valley12, kioseoglou:221907,Lagarde14,Heinz14}. The theory of spin and valley relaxation mechanisms in MoS$_2$ is currently under intense investigation~\cite{Dery_Relax,Ochoa13,MWWu14, Glazov_14,MWWu_arxiv}.  The combination of 2D confinement, weak dielectric screening and large effective masses of electrons and holes also leads to very large excitonic effects in single-layer dichalcogenides~\cite{Lambrecht12,Komsa12,Heinz_Trions,Wirtz13,Malic14}. Recently, the carrier~\cite{Korn_APL11,Zhao_12}  and valley~\cite{Wang13,Lagarde14,CongMai14} dynamics in MoS$_2$ have been studied by several groups, either with time-resolved PL~\cite{Korn_APL11,Lagarde14}, or by utilizing  two-color pump-probe~\cite{Zhao_12,Wang13,CongMai14} schemes.

Here, we present time-resolved pump-probe experiments in which we study the carrier and valley dynamics in a mechanically exfoliated single-layer MoS$_2$ flake at low temperatures. We utilize time-resolved Kerr rotation (TRKR), a well-established technique~\cite{awschalom94_2} for studying electron and hole spin dynamics in semiconductor heterostructures~\cite{KornReview}, as well as transient reflectivity ($\Delta R$) measurements. Both techniques yield sub-ps time resolution and allow resonant excitation of our sample.  We demonstrate that for resonant excitation conditions, the valley lifetime exceeds the photocarrier lifetime, corresponding to a transfer of optically generated valley polarization to resident carriers (electrons). For nonresonant excitation, the valley lifetime decreases while the photocarrier lifetime increases.

MoS$_2$ flakes are prepared from mineral bulk MoS$_2$ crystals using a recently developed all-dry transfer technique~\cite{Gomez_Transfer}.  Hereby,  MoS$_2$ is at first exfoliated onto a transparent, viscoelastic Polydimethylsiloxane (PDMS) film, which is attached to a glass slide, using adhesive tape. The optical contrast in a  microscope is sufficient to identify monolayers of MoS$_2$ on top of the PDMS, and selected flakes can be transferred onto other substrates. We utilize a  p-doped  silicon wafer with 300~nm SiO$_2$ layer and lithographically defined metal markers as final substrate. Working under an optical microscope, the glass slide is turned upside down and clamped into a xyz stage. The PDMS film, which faces  downwards, is lowered by the stage and brought into contact with the SiO$_2$ substrate. By carefully lifting the microscope slide again, we are able to transfer the MoS$_2$ flake from the PDMS film onto a pre-determined position on the SiO$_2$. This method yields substantially larger single-layer flakes than direct mechanical exfoliation of MoS$_2$ onto SiO$_2$ using adhesive tape.

Characterization measurements are carried out using Raman spectroscopy and PL measurements at room temperature.  Details of the experiments and data treatment are published elsewhere~\cite{Plechinger14}.

All time-resolved measurements are performed using a tunable, frequency-doubled pulsed fiber laser system (pulse length 150~fs, spectral linewidth 6~meV). The laser beam is split into  pump and probe pulse trains, and a variable delay between pulses is realized using a mechanical delay stage. Pump and probe beams are focussed onto the sample  mounted in the He-flow cryostat, using an achromatic lens to a spot size of about 80~$\mu$m, so that the pump beam is at normal incidence, while the probe beam has an angle of about 4 degrees to the sample normal. This allows us to spatially separate the reflected probe beam from the pump beam. An excitation density of about 30~Wcm$^{-2}$ is used in the pump beam, the probe beam has a similar intensity. If we assume an absorption efficiency of 10~percent in the MoS$2$ layer for resonant excitation, this excitation density corresponds to an optically generated carrier density of 8$\times$10$^{10}$~cm$^{-2}$, which is 1-2 orders of magnitude lower than the residual n-doping typically observed in exfoliated MoS$_2$ flakes~\cite{Novoselov26072005,Kis_NatNano11}. A digital microscope system allows to position the MoS$_2$ flake in the laser focus. A mechanical chopper is used to modulate the pump beam intensity, allowing for lock-in detection of the pump-induced changes in probe beam intensity and polarization. In $\Delta R$ measurements, both beams are linearly polarized orthogonal to each other, and the reflected probe beam intensity is detected as a function of pump-probe delay using a photodiode. In TRKR measurements, the pump beam is circularly polarized using an achromatic quarter-wave plate, while the probe beam is linearly polarized. To detect changes of the reflected probe beam polarization state, an optical bridge detector is employed.

\begin{figure}
\includegraphics*[width=\linewidth]{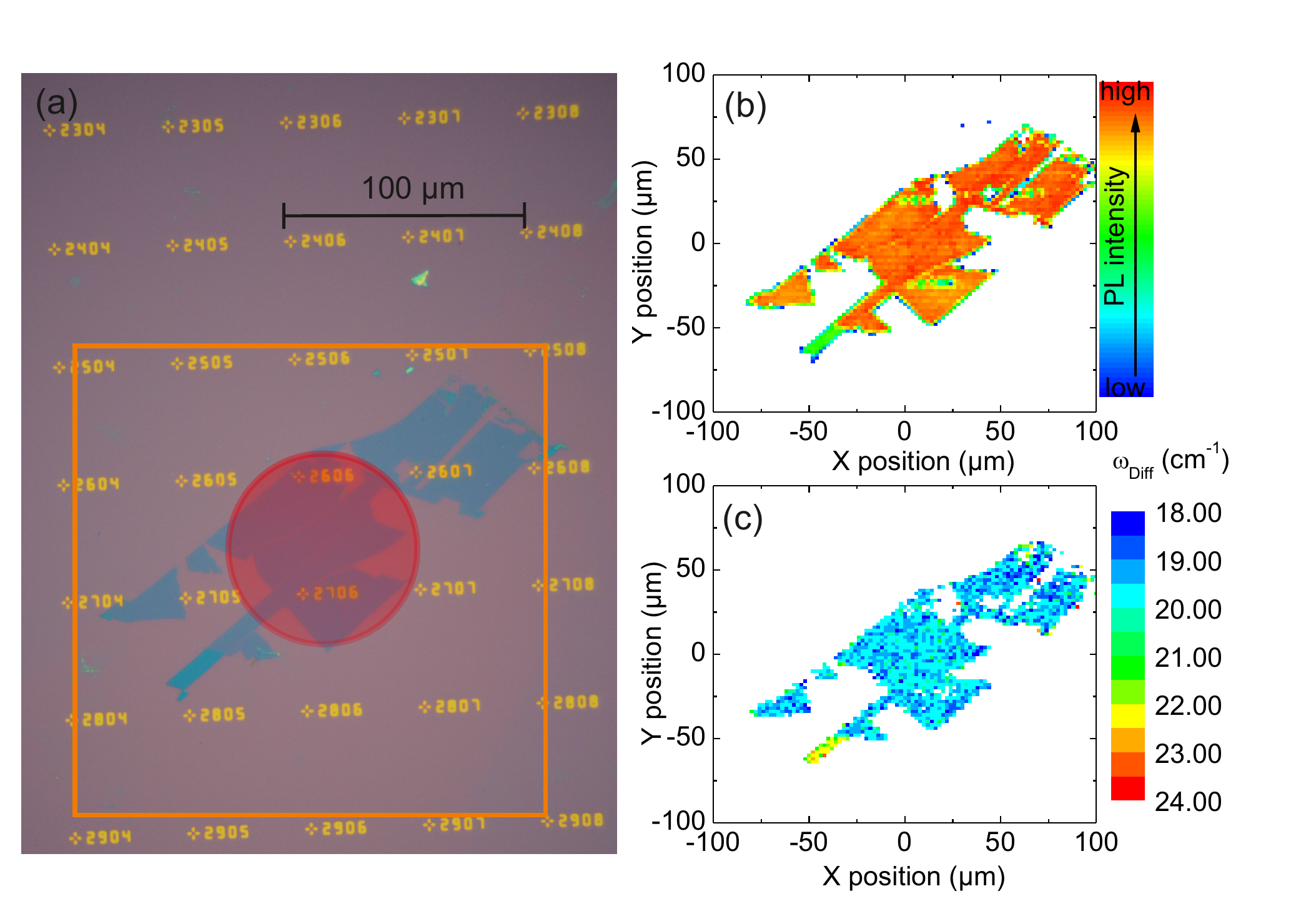}
\caption{(a) Optical micrograph of MoS$_2$ flake. The orange square indicates the scan area for the false color maps shown in (b) and (c), the semi-transparent circle indicates the focal spot size for the pump-probe measurements. (b) False color map of A exciton PL intensity as a function of position. (c) False color map  of the difference between the A$_{1g}$ and E$^1_{2g}$ mode frequencies.}
\label{Fig1}
\end{figure}

First, we discuss characterization measurements of our MoS$_2$ flake. Figure~\ref{Fig1}(a) shows an optical micrograph of the flake on the Si/SiO$_2$ substrate. A large optical contrast is observed on such substrates even for single-layer MoS$_2$ flakes due to interference effects~\cite{castellanos-gomez:213116,Kis11}.  To quantify the number of layers, we utilize scanning Raman and PL measurements. A false color map of the integrated intensity of the A exciton PL is depicted in Fig.~\ref{Fig1}(b). We clearly see a near-homogeneous intensity of the PL emission throughout most of the flake, with a significantly weaker (by a factor of 3) intensity observed only in the lower left part. This indicates that the majority of the flake consists of a single MoS$_2$ layer, while the lower-left part is a bilayer. A similar reduction of the PL yield for the single-layer-bilayer transition was reported by several groups~\cite{Splen_Nano10,Maultzsch12}, while Mak et al. observed an even larger reduction~\cite{Heinz_PRL10}. We note that the A exciton peak observed in the PL spectra actually consists of, both, emission from neutral and negatively charged excitons (trions, A$^-$)~\cite{Heinz_Trions}, as MoS$_2$ flakes prepared from mineral bulk material have a residual n-doping, which is also observed in transport experiments~\cite{Novoselov26072005,Kis_NatNano11}. Due to the large PL linewidth, the energy splitting between A and  A$^-$ cannot be resolved in our measurements.

Raman measurements confirm our interpretations. In MoS$_2$, the frequencies of the most prominent Raman modes A$_{1g}$ and E$^1_{2g}$~\cite{Heinz_ACSNano10,Molina11}, as well as those of the low-energy shear and compression modes~\cite{plechinger:101906,Zeng_ShearMos,Ferrari_ShearMos,Dresselhaus_Shear}, strongly depend on the number of layers.  In Fig.~\ref{Fig1}(c), we plot the (A$_{1g}$ - E$^1_{2g}$) frequency difference as a function of position on the flake in a false color map. Again, we find a near-homogeneous contrast, corresponding to a frequency difference of about 19~cm$^{-1}$, in good agreement with values reported for monolayers~\cite{Heinz_ACSNano10}, for the majority of the flake. Only the lower left region shows a larger frequency difference of about  21~cm$^{-1}$, corresponding to a bilayer.  In low-temperature PL measurements using near-resonant, circularly polarized excitation, we find a circular polarization degree of the A exciton PL of about 50~percent. To summarize our characterization measurements, we find that our MoS$_2$ flake consists of a homogeneous large-area monolayer region and shows strong valley polarization at low temperatures. Thus, it is ideally suited to study valley dynamics using a large focal spot. (See supplementary material for Raman and PL spectra used in sample characterization.)
\begin{figure}
\includegraphics*[width=\linewidth]{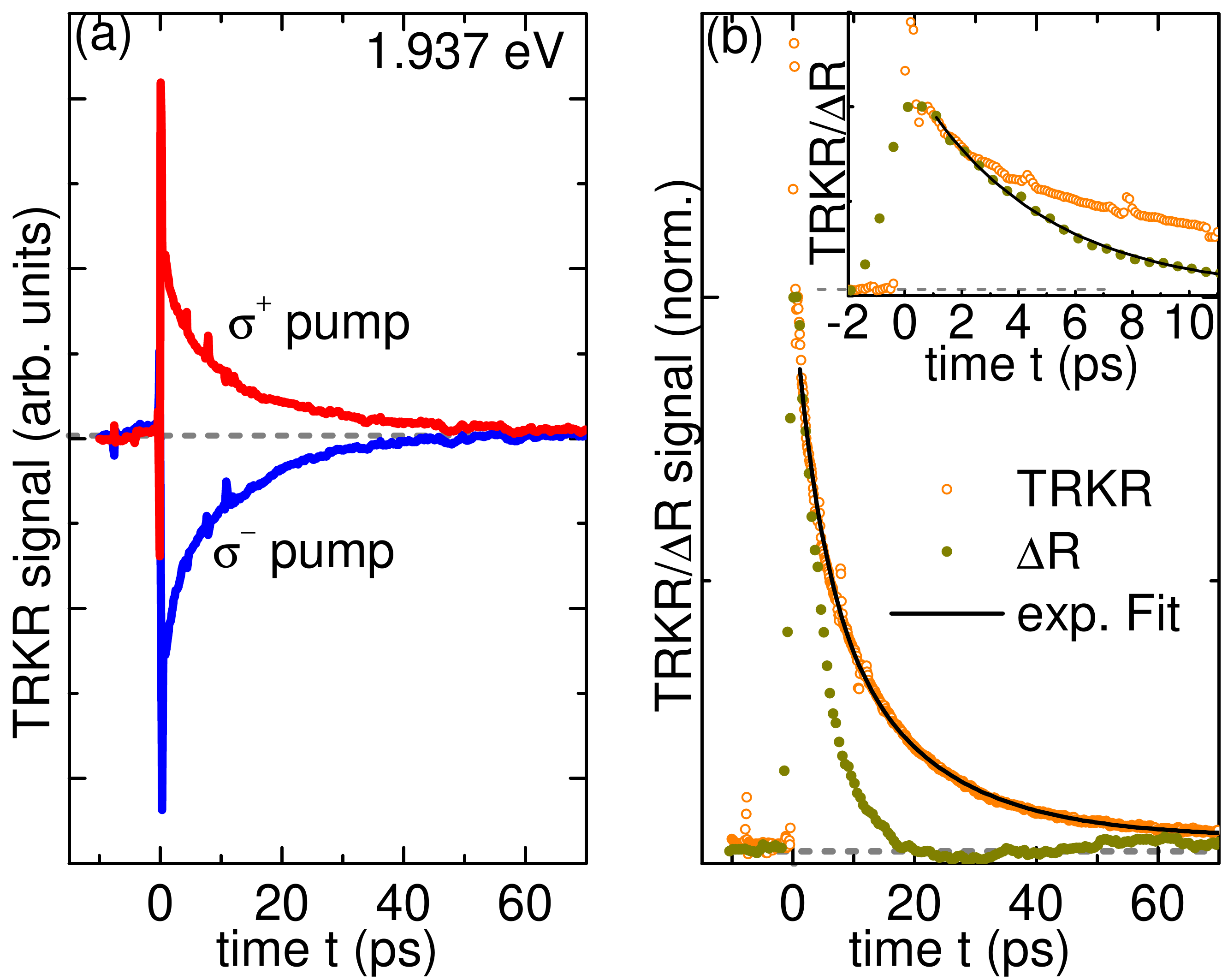}
\caption{(a) TRKR traces measured on the MoS$_2$ flake  for different helicities of the pump beam, with a laser excitation energy of 1.937~eV. (b) Normalized TRKR and transient reflectivity traces measured on the MoS$_2$ flake with a laser excitation energy of 1.937~eV. The solid line indicates a biexponential fit to the TRKR trace. The inset shows a high-resolution plot of the data, The solid line indicates a monoexponential fit to the transient reflectivity trace.}
\label{Fig3}
\end{figure}
\begin{figure}
\includegraphics*[width=\linewidth]{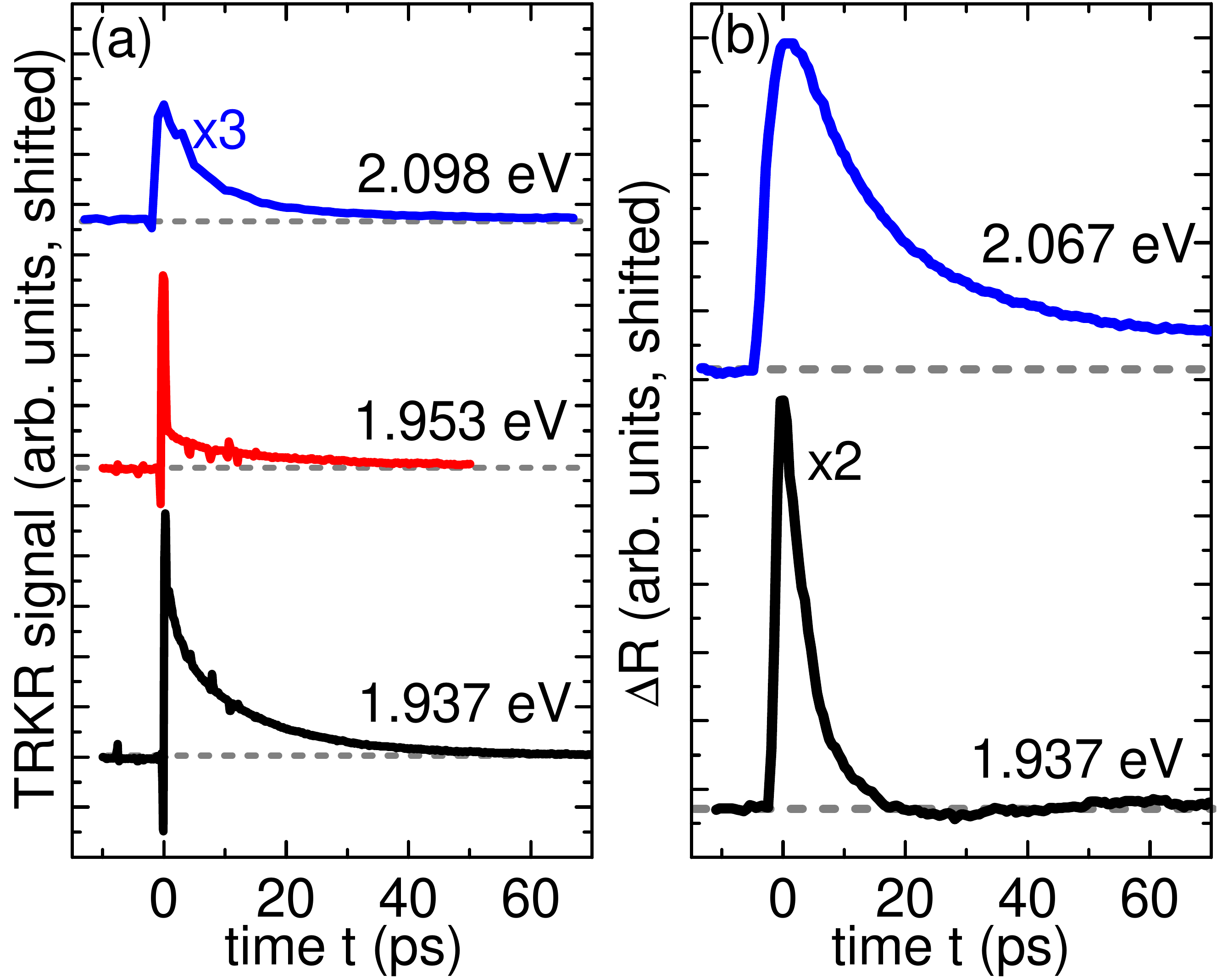}
\caption{(a) TRKR traces for different laser excitation energies. (b) Transient reflectivity measurements on the MoS$_2$ flake for different laser excitation energies.}
\label{Fig4}
\end{figure}

Now, we discuss the time-resolved measurements. All measurements shown here were performed at a sample temperature of 4.5~K. Figure~\ref{Fig3}(a) shows two TRKR traces measured using resonant excitation. We clearly see that the TRKR signal flips its sign as the pump beam helicity is changed from $\sigma^+$ to $\sigma^-$. By contrast, for linearly polarized excitation, we do not observe a pronounced TRKR signal (See supplementary material for comparison of TRKR traces and discussion of data treatment). This clearly indicates that the TRKR signal indeed stems from the occupation imbalance of the K$^+$/K$^-$ valleys.  In Fig.~\ref{Fig3}(b), we compare TRKR and $\Delta R$ traces for resonant excitation directly. We find that the $\Delta R$ signal shows a monoexponential decay with a decay constant of about 4.5~ps, and we may identify this as the photocarrier recombination time $\tau_C$, in agreement with low-temperature TRPL experiments~\cite{Korn_APL11,Lagarde14}. By contrast, the TRKR trace shows a clear biexponential decay, which we fit with a fixed value $\tau_C$=4.5~ps for the fast decay.
We find a second, slow decay constant $\tau_V$=17~ps. We note that, in contrast to helicity-resolved TRPL experiments, which measure the  polarization \emph{degree} of the PL emission, the Kerr rotation angle is proportional to the spin/valley polarization, i.e., the absolute number of polarized carriers in the area/volume tested by the probe laser. In the absence of spin or valley relaxation processes, the polarization degree in TRPL experiments would therefore remain constant throughout the photocarrier lifetime, while the Kerr rotation signal would decay, as the number of polarized carriers is reduced by photocarrier recombination. Thus, we can associate the fast decay component in the TRKR trace with photocarrier recombination.
The fact that a finite TRKR signal is still observable at time delays exceeding the photocarrier lifetime indicates that a partial transfer of spin and valley polarization from optically oriented electron-hole-pairs to resident electrons occurs. Thus, the  polarization decay time extracted from our data is a measurement of spin/valley dynamics for \emph{resident} electrons in exfoliated MoS$_2$ at liquid-helium temperatures. It is in reasonable agreement with recent theory calculations~\cite{MWWu14, Glazov_14}.
We now turn to nonresonant excitation experiments. As we increase the laser energy by about 16~meV (middle trace in Fig.~\ref{Fig4}(a)), we find that the shape of the TRKR trace changes drastically, about 80~percent of the initial TRKR signal decays within about 600~fs, while the remaining signal decays on the 20~ps timescale. We may tentatively identify the rapid initial decay as a partial loss of carrier spin or valley polarization during energy relaxation. Additionally, we need to consider our degenerate measurement scheme, with  identical pump and probe beam energies: the Kerr rotation angle for a given  polarization sensitively depends on the probe energy, and is strongly enhanced close to a resonance in the material. Thus, by detuning the laser, we also decrease our detection sensitivity for spin- and valley-polarized carriers at the band extrema, and a part of the rapid initial signal decay may be related to pure energy relaxation of valley-polarized carriers. As we increase the laser energy further, the amplitude of the TRKR signal  decreases, and the signal is lost for laser energies above 1.99~eV (see  supplementary material for the energy dependence of the Kerr signal amplitude). For a laser energy that matches the B exciton resonance, we find a weak TRKR signal again, which shows a monoexponential decay with a decay constant of about 7~ps (top trace in Fig.~\ref{Fig4}(a)). This indicates that our TRKR technique is also suitable to study valley dynamics of B excitons, which are difficult to observe in PL-based measurements. We can directly compare the valley dynamics measured by TRKR to the carrier dynamics accessible in $\Delta R$ traces, as depicted in  Fig.~\ref{Fig4}(b). For resonant excitation, we find the rapid, monoexponential decay of the transient reflectivity discussed above. When we increase the laser energy by more than 30~meV, we find longer-lived $\Delta R$ traces with decay constants of about 15~ps, almost independent of the laser energy in the investigated range up to 2.07~eV. An increase of the photocarrier lifetime under nonresonant excitation is a common phenomenon in direct-gap semiconductors, as excitons generated with excess energy first have to relax their center-of-mass momentum before they may recombine radiatively~\cite{Shields00,Zhukov07}. As discussed above, we note that due to our degenerate pump-probe scheme, $\Delta R$ traces measured for nonresonant excitation may include both, decay processes due to photocarrier recombination and interband energy relaxation. Thus, the $\Delta R$ decay observed under these conditions is only a lower bound for the photocarrier recombination time. Our observations are in good agreement with previous studies of the excitation-energy-dependent PL circular polarization degree~\cite{kioseoglou:221907}: for slightly nonresonant excitation, a partial loss of valley polarization occurs on the sub-ps timescale, while the remaining valley polarization decays with a rate commensurate to the photocarrier lifetime, so that circularly polarized PL is observed. For larger excess energy, even when the B exciton resonance is addressed, the valley polarization fully decays significantly faster than the photocarriers, so that no circular polarization of the PL can be detected.

In conclusion, we have demonstrated time-resolved Kerr rotation measurements on exfoliated monolayer MoS$_2$ at low temperatures. Compared to time-resolved photoluminescence, this technique yields a higher temporal resolution and allows for resonant excitation of a sample. Additionally, it enables us to study the valley dynamics of \emph{resident} carriers in doped MoS$_2$.
The authors  gratefully acknowledge fruitful discussion with A. Castellanos-Gomez, A. Molina-Sanchez and L. Wirtz, as well as financial support by the DFG via SFB689, KO3612/1-1 and GRK 1570.

\newpage
\section{Supplementary information}
\subsection{Sample characterization by photoluminescence and Raman spectroscopy}
\begin{figure}
\includegraphics*[width=\linewidth]{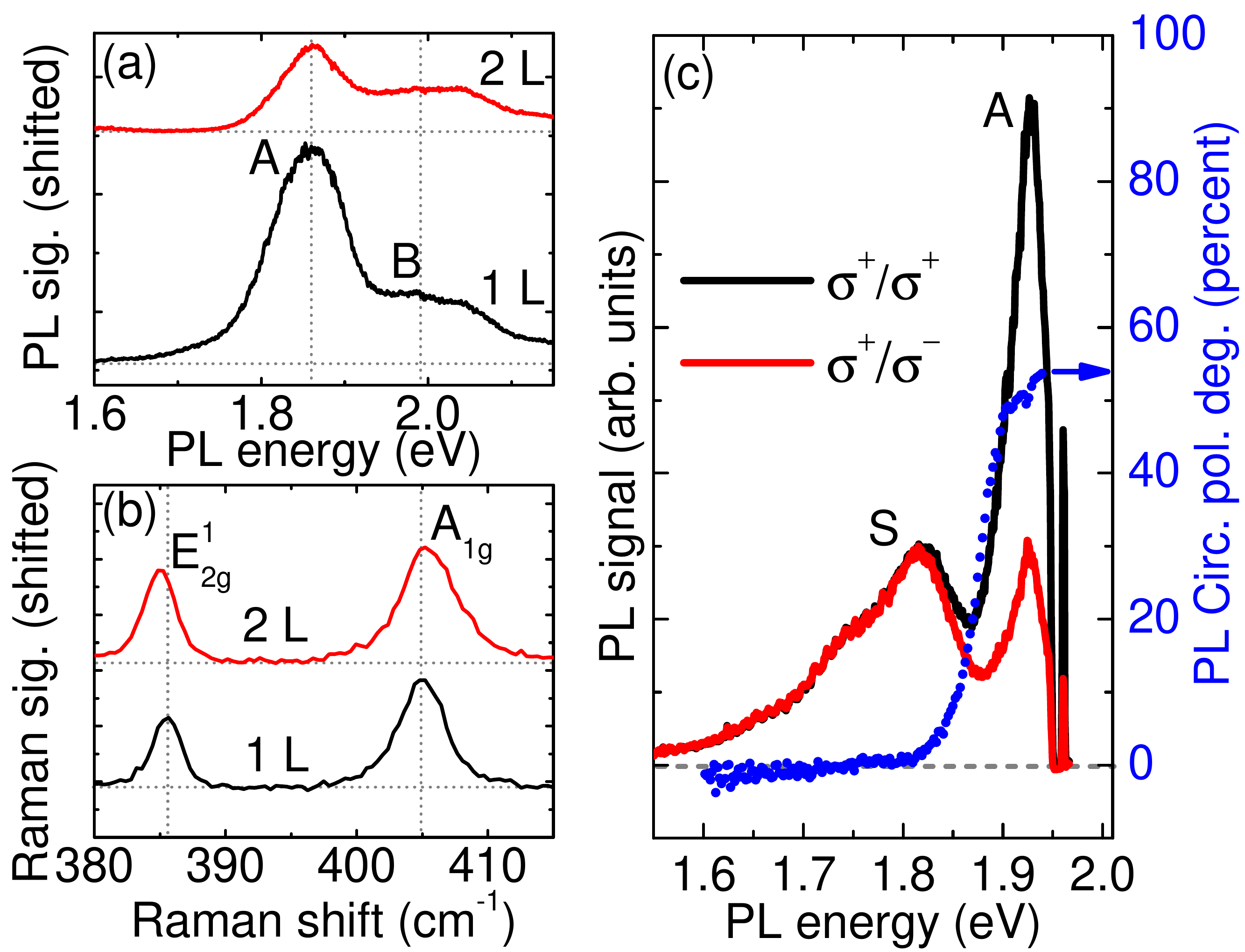}
\caption{ (a) Room-temperature PL spectra collected on single-layer (1L) and bilayer (2L) part of the flake. (b) Room-temperature Raman spectra collected on single-layer and bilayer parts of the flake. (c) Solid lines: Helicity-resolved PL spectra collected under near-resonant excitation at 4.5~K. Circular dots: circular polarization degree of the PL extracted from the spectra.}
\label{FigS1}
\end{figure}

PL and Raman measurements for characterization of the flake are performed at room temperature using an excitation density of 220~kWcm$^{-2}$. The false color maps depicted in the main manuscript are created from spectra collected on a square lattice with a step size of 2~$\mu m$. Figure~\ref{FigS1} (a) shows two characteristic spectra collected on the single- and bilayer parts of the flake. They are generated by averaging over 3 by 3 individual spectra collected in the scanning PL measurement to increase the signal/noise ratio. We clearly see a well-defined A exciton peak at about 1.86~eV, and a broader B exciton peak at about 2~eV, in both the single- and the bilayer parts of the flake, in agreement with experimental observations by Splendiani et al.~\cite{Splen_Nano10}. The PL intensity of the A exciton emission is about 3 times weaker in the bilayer part of the flake.

In Fig.~\ref{FigS1} (b), Raman spectra for single- and bilayer parts of the flake are compared directly. Again, averaging over 3 by 3 individual spectra collected in the scanning Raman experiment is performed to increase the signal/noise ratio. The Raman spectrum for the singlelayer part of the flake shows a frequency difference (A$_{1g}$ - E$^1_{2g}$) of 19~cm$^{-1}$.  In the bilayer part, the A$_{1g}$ mode blueshifts, while the E$^1_{2g}$ mode redshifts, so that the frequency difference increases to 21~cm$^{-1}$, in agreement with previous reports~\cite{Heinz_ACSNano10}.

For low-temperature PL measurements, the samples are mounted in vacuum on the cold finger of  a small He-flow cryostat, which can also be scanned under the microscope.   To study valley polarization effects, near-resonant excitation is employed in the PL setup. For this, a 633~nm cw He-Ne laser is used. This laser is circularly polarized by a quarter-wave plate and coupled into the microscope system. The circular polarization of the PL is analyzed using a second quarter-wave plate and a linear polarizer,  and a longpass filter placed in front of the spectrometer is utilized to suppress scattered laser light. For these measurements, we focus the near-resonant excitation laser onto the monolayer part of the flake. Figure~\ref{FigS1}(c) shows two PL spectra measured at a sample temperature of 4.5~K on the same sample spot for co- and contracircular helicities of excitation and detection, respectively. At low temperatures, besides the A/A$^-$ exciton emission, a second, lower-energy peak (S exciton) is observable in MoS$_2$~\cite{Korn_APL11}. It can be associated to excitons bound to surface adsorbates~\cite{Plechinger12}, and is suppressed with increasing temperature. In the helicity-resolved spectra, we find a significantly larger A exciton PL emission for co-circular helicity, indicating a pronounced valley polarization. By contrast, the S exciton emission is nearly identical for both spectra. From the two spectra, the circular polarization degree of the PL is calculated by dividing the difference of the PL intensities by their sum. For this, the PL intensities are spectrally averaged in  2.5~meV wide windows to reduce the noise. We find a circular polarization degree of more than 50~percent for the A exciton emission, while the S exciton emission is unpolarized. Similar, and even larger values of the A exciton polarization have been found by several groups~\cite{Yao12,Heinz12,Zeng_WS2_12, Marie_Valley12, Lagarde14}, and a strong dependence on the excitation energy was reported recently~\cite{kioseoglou:221907}.
\subsection{Effects of pump beam polarization on TRKR signal}
\begin{figure}
\includegraphics*[width=\linewidth]{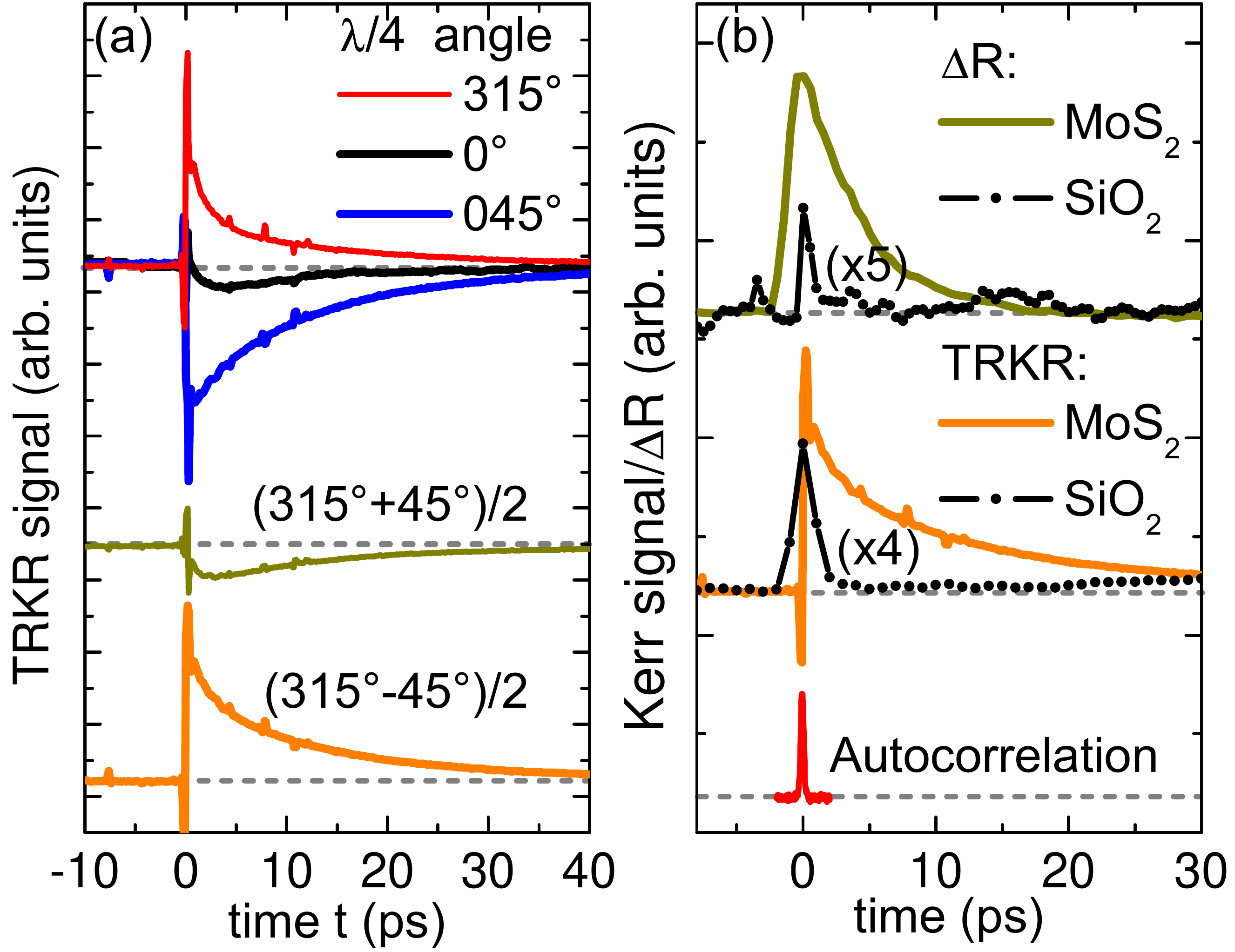}
\caption{ (a) TRKR signals for different helicities of the pump beam and linearly polarized excitation, compared to the difference and sum signals of the different helicities.   (b) $\Delta$R and TRKR signals on the MoS$_2$ and the bare SiO$_2$/Si wafer.}
\label{FigS2}
\end{figure}

In order to clearly identify the origin of the TRKR signal, we perform a series of TRKR measurements for different polarization states of the pump beam. Figure~\ref{FigS2}(a) shows three TRKR traces measured under resonant excitation conditions with constant excitation density. We utilize a combination of a linear polarizer and an achromatic quarter-wave plate to change the pump beam helicity. If the fast axis of the quarter-wave plate is aligned parallel (at an angle of zero degrees) to the linear polarizer, linear polarization of the pump beam results, for angles of 315 or 45 degrees, left- or right-handed circular polarization of the pump beam is generated. For circularly polarized excitation, we find pronounced TRKR signals, and the sign of the Kerr signal flips as we flip the pump beam helicity. However, for linearly polarized excitation, we also observe a  TRKR signal, which is significantly weaker than for circularly polarized excitation. Most likely, the linear polarization of the pump beam becomes slightly elliptical as it passes through the cryostat window due to strain-induced birefringence in the glass, leading to a small valley polarization and a subsequent Kerr signal. Close examination of the TRKR traces under circularly polarized excitation also reveals slightly different signal shapes, indicating an imbalance between left- and right handed excitation. Remarkably, the sum of the TRKR signals for the two helicities closely matches the TRKR signal recorded for linearly polarized excitation, supporting our interpretation that there is a slight distortion of the polarization state of the pump beam. Consequently, by taking the difference of the TRKR signals for the two helicities, any signals that do not depend on helicity are suppressed. All TRKR traces depicted in the main manuscript, with the exception of those shown in Fig. 2(a), are generated by taking the difference of TRKR signals with $\sigma^+$ and $\sigma^-$ excitation. The two traces shown in Fig. 2(a) are generated by subtracting the TRKR signal measured for linearly polarized pump from the signals measured with $\sigma^+$ or $\sigma^-$ excitation, respectively.
\subsection{Control experiments}
To prove that the $\Delta$R and TRKR signals we observe are caused by the MoS$_2$, we perform control experiments on the bare SiO$_2$/Si wafer next to the MoS$_2$ flake. As Fig.~\ref{FigS2}(b) shows, we only observe very weak $\Delta$R and TRKR signals on the bare substrate, which decay on the picosecond timescale.

\subsection{Excitation-energy dependence of TRKR signal}
\begin{figure}
\includegraphics*[width=0.8 \linewidth]{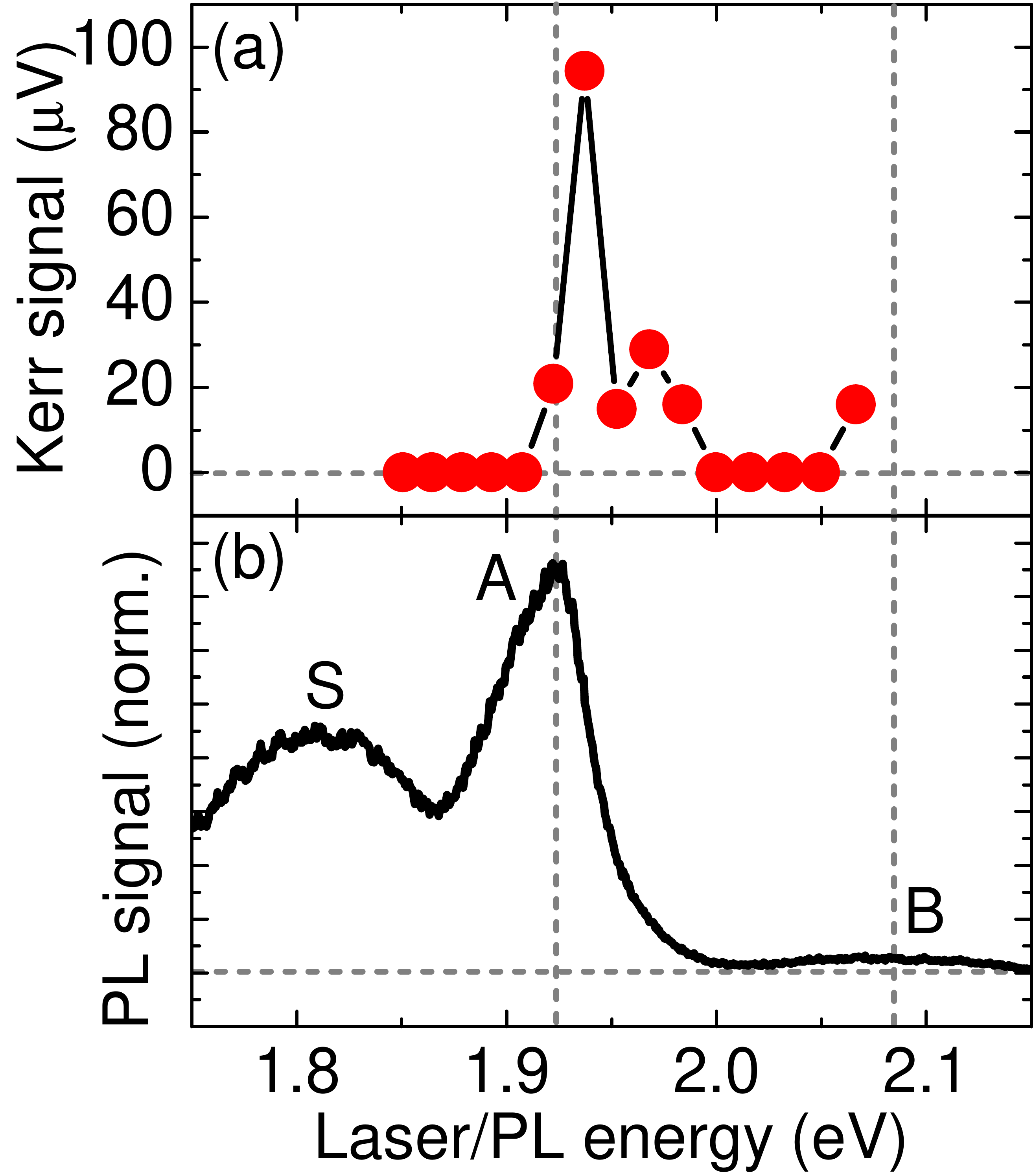}
\caption{ (a) Kerr signal amplitude at a fixed delay position 2.5~ps after arrival of the pump pulse. (b) PL spectrum recorded at 4.5~K using 532~nm laser excitation.}
\label{FigS3}
\end{figure}

A series of TRKR measurements as a function of the laser excitation energy is performed at a sample temperature of 4.5~K. From these measurements, we extract the amplitude of the Kerr signal at a fixed time delay between pump and probe pulses of 2.5~ps. We note that as we tune the laser excitation energy, the total laser power varies by about 25~percent throughout the energy range investigated. To account for these changes, we scale the extracted Kerr amplitude for each excitation energy to the laser power. This corrected amplitude is depicted in Fig.~\ref{FigS3}(a). We clearly see a pronounced maximum at a laser energy of 1.937~eV. As the laser energy is increased, the Kerr signal amplitude drops, and between 1.99~eV and 2.05~eV, no TRKR signal can be observed. A small TRKR signal is recovered for larger excitation energies close to the B exciton resonance. We can directly compare the energy dependence of the TRKR amplitude to a PL spectrum collected at 4.5~K using 532~nm laser excitation, shown in Fig.~\ref{FigS3}(b). We find that the maximum Kerr signal amplitude is observed for energies slightly above the A exciton peak emission in photoluminescence, indicating a small Stokes shift of about 10~meV.

\end{document}